\documentclass[a4paper]{article}

\usepackage{INTERSPEECH2021}

\usepackage{algorithm}
\usepackage{algpseudocode}
\usepackage{multirow}
\usepackage{caption}

\title{Text-to-speech for the hearing impaired}
\name{Josef Schlittenlacher$^1$, Thomas Baer$^2$}
\address{
  $^1$University of Manchester\\
  $^2$University of Cambridge}
\email{josef.schlittenlacher@manchester.ac.uk, tb107@cam.ac.uk}

\begin{document}

\maketitle
\begin{abstract}
Text-to-speech (TTS) systems offer the opportunity to compensate for a hearing loss at the source rather than correcting for it at the receiving end. This removes limitations such as time constraints for algorithms that amplify a sound in a hearing aid and can lead to higher speech quality. We propose an algorithm that restores loudness to normal perception at a high resolution in time, frequency and level, and embed it in a TTS system that uses Tacotron2 and WaveGlow to produce individually amplified speech. Subjective evaluations of speech quality showed that the proposed algorithm led to high-quality audio with sound quality similar to original or linearly amplified speech but considerably higher speech intelligibility in noise. Transfer learning led to a quick adaptation of the produced spectra from original speech to individually amplified speech, resulted in high speech quality and intelligibility, and thus gives us a way to train an individual TTS system efficiently.
\end{abstract}

\section{Introduction}
\label{sec:intro}

About 20\% of all adults have a hearing loss~\cite{WHO2021,Blackwell2014,Dillon2020} but only a minority of them have access to hearing aids. Even among those who own hearing aids, about 50\% use them rarely or not at all~\cite{Dillon2020} although an effective amplification reduces the handicap even for mild to moderate hearing losses. Several of the reasons for not wearing a hearing aid are related to the device, like comfort, maintenance, appearance, psycho-social factors, dissatisfying sound quality or financial reasons~\cite{Mccormack2013}.

Conversational artificial intelligence, or more general technical devices that produce synthesized speech, afford the opportunity to communicate to hearing-impaired people without the need of hearing aids by shifting the amplification from the receiver to the source and directly delivering the desired amplified speech.

Recent advances in text-to-speech (TTS) systems have increased the sound quality of generated speech compared to traditional approaches like the Griffin-Lim algorithm~\cite{Griffin1984} that estimates the phase information for a magnitude spectrogram in an iterative procedure. The first neural vocoder was the WaveNet~\cite{Oord2016}, which directly generates the waveform using an autoregressive neural network. Because the generation of one sample is based on all past samples, including the immediately preceding sample, it is rather slow at inference time. Several approaches have overcome this limitation and presented neural vocoders that can run faster than real time on a modern graphics processing unit. Examples are WaveGlow~\cite{Prenger2019}, SampleRNN~\cite{Mehri2017}, Parallel WaveNet~\cite{Oord2018}, ClariNet~\cite{Ping2018}, WaveRNN~\cite{Kalchbrenner2018}, WaveGAN~\cite{Donahue2019}, FloWaveNet~\cite{Kim2019}, LPCNet~\cite{Valin2019}, MelNet~\cite{Vasquez2019}, SqueezeWave~\cite{Zhai2020} or WaveGrad~\cite{Chen2020}. Most of them are not directly conditioned on text but on Mel spectrograms, which can be predicted for a desired text by systems such as Tacotron2~\cite{Shen2018}, Deep Voice 3~\cite{Ping2017}, FastSpeech~\cite{Ren2019} or Flowtron~\cite{Valle2020}.

Instead of neural vocoders being trained on the recorded speech, they could also be trained to generate speech that is amplified to compensate for the hearing loss of a given listener. Training TTS for different targets was done for other purposes such as to generate speech for speech-impaired listeners with a parametric TTS system~\cite{Jreige2009} or to enhance speech in noise by using the Lombard effect~\cite{Paul2020}.

Producing the desired sound at the source has considerable advantages compared to additional processing in a hearing aid: Hearing-aid algorithms must not introduce a delay greater than 6 ms in order to not cause distortions due to the time difference between the direct sound that bypasses the hearing aid and the amplified sound that the hearing aid produces~\cite{Stone2008}. By contrast, the speech for the training set of a neural vocoder can be processed using an arbitrary portion in the time domain. In theory it could use the whole sound already for the first samples.

Hearing-aid algorithms aim to optimize speech intelligibility~\cite{Launer2016} and to restore loudness to normal~\cite{Launer2003}. They typically do so by dividing the spectrum into about 10 to 20 channels, and calculate a level-dependent gain for each channel. The parameters are obtained from the audiogram by a “prescription formula”~\cite{Moore2010,Keidser2011}. That gain is then applied between a Fourier Transform and its inverse, or to adjust the gains of a bandpass-filtered signals before they are added back together. A quick change in the gain can lead to audible artefacts or glitches. This could be overcome for the training set of a neural vocoder by looking ahead in time and by choosing a very small hop size between analysis windows so that the gain is always accurate and changes smoothly between any two samples of the processed speech.

A drawback of a neural vocoder that directly generates amplified speech is that it needs to be trained for each specific hearing impairment. Transfer learning can alleviate this problem and reduce the training time significantly by taking a trained model for normal hearing as a starting point. A similar strategy was successful for enhancing speech in noise~\cite{Paul2020}.

In this paper we propose to train a TTS system to directly generate individually amplified speech from text (IA-TTS), and present an algorithm to restore sounds for hearing-impaired listening to normal loudness that is not limited to the constraints of processing sounds in a hearing aid. Our TTS system is similar to the work of Paul et al.~\cite{Paul2020} but for the purpose of compensating for a hearing loss rather than enhancing speech in noise. The amplification algorithm shares the goals with that of Launer and Moore~\cite{Launer2003} but focused on speech quality rather than circumventing the limitations of processing in a hearing aid. In particular, it uses a hop size of only one sample. We tested the system using Tacotron2 and a WaveGlow that was trained on the individually amplified targets. Subjective evaluations were obtained from the hearing-impaired listener for which the system was trained, and after adding the inverse amplification also by normal-hearing listeners. The possibility of reducing training time for an individual hearing impairment by using transfer learning was evaluated.

\section{Model}
\label{sec:model}


\subsection{TTS system and training}
\label{ssec:tts}

The primary purpose of our TTS model is to generate speech that is already amplified to compensate for a hearing loss via a neural vocoder. We trained it on the individually amplified speech targets, so that the algorithm for amplifying the sounds is only required during training stage and the model complexity is the same as that of a normal TTS system during inference. Doing so has several major advantages: In contrast to applying a hearing-aid algorithm after synthesizing speech for normal hearing, no delay is added. This allows us to use an arbitrarily long window size in the time domain, and even to consider the future signal in the amplification algorithm. Furthermore, the amplification algorithm is not bound to real-time processing constraints since it does not need to be run at inference time and the training set only needs to be processed once.

For converting text to Mel spectrograms we used a pretrained Tacotron2, which is a recurrent sequence-to-sequence network with an attention mechanism. We used WaveGlow as a neural vocoder to transform Mel spectrograms to speech. WaveGlow is a flow-based~\cite{Rezende2015} generative neural network that transforms a simple Gaussian distribution to a desired more complex distribution, most commonly speech but it could also be that of other sounds like music~\cite{Zhao2020}. We conditioned it on Mel spectrograms of speech for normal hearing, i.e. the same as in the original model, but trained it to generate speech that was processed through an amplification algorithm.

The amplification algorithm is a frequency-level dependent amplification that is based on perceptual mechanisms such as auditory filters and loudness to compensate for a hearing loss. It is explained in detail in the next section. We used the LJSpeech corpus~\cite{Ljspeech17}, which has a sample rate of 22050 Hz, to derive Mel spectrograms and the amplified speech.

\subsection{Hearing-loss compensation}
\label{ssec:amp}

Restoring a sound to normal loudness for a hearing-impaired listener needs to take into account the hearing loss at a particular frequency. This is commonly measured by an audiogram, which gives the amount of hearing loss at threshold but not the type of hearing loss. For the most common type of hearing loss, a cochlear hearing loss, loudness perception also depends on sound pressure level: While a large amplification is necessary to restore soft loudness levels, loudness perception is close to normal at high levels~\cite{Moore2004}. For this purpose hearing aids typically apply a linear gain at low levels, and a compressed gain at high levels~\cite{Launer2016}. We decided not use such a parametric approach but to determine the exact gain throughout the level range.

Determining the necessary compensation is further complicated because loudness does not only depend on a single frequency bin, which would be a single point on the basilar membrane, but on the intensity that falls within the auditory filter around this frequency. We calculate auditory filters on the ERB-scale~\cite{Glasberg1990}.

In our algorithm, the gain for each frame and frequency bin is determined as follows: (1) The level that falls in the auditory filter centered at that frequency bin is calculated. (2) The specific loudness of a uniformly exciting noise with that auditory-filter level is calculated for normal hearing. A uniformly exciting noise is a noise with equal intensity in each auditory filter. (3) The auditory-filter level of a uniformly exciting noise that yields the same specific loudness at that frequency for the impaired hearing is determined. (4) The gain for the frequency bin is the level difference between steps 3 and 2. Step 3 needs to be done in an iterative search but steps 2 to 4 can be stored in a look-up table. The loudness calculations in steps 2 and 3 were done using the Moore-Glasberg loudness model for hearing-impaired listening~\cite{Moore2004}, which also includes normal hearing as a special case (0 dB hearing loss).

The generation of individually amplified speech from normal speech using this procedure for a single frame is described in Algorithm~\ref{alg1}. It is based on windows of 1024 samples for which the amplification is obtained and that are processed at a hop size of only 1 sample. From each processed window, only the central sample is taken. Taking only one sample is equivalent to a very narrow window in the time domain, which affects all frequencies equally. Because windows are moved by only one sample, which corresponds to 0.05 ms, and also produce only a single sample, the outputs can easily be concatenated. Most importantly, there are no large changes in amplification or phase between two subsequent frames. We think that this seamless concatenation of consecutive frames, and the detailed loudness reconstruction, can mitigate the loss in sound quality that is usually observed for hearing-aid algorithms. The computation time on an i7 processor was about 3 days for the LJSpeech corpus, 3 times slower than real time.

\begin{algorithm}
\caption{Individual amplification to compensate for a hearing loss}\label{alg1}
\begin{algorithmic}[1]
\Repeat
\Comment{hop size 1 sample}
\State Take 1024-point DFT
\State Calculate auditory-filter levels for each bin
\State Look-up and apply gain for each bin
\State Take inverse DFT
\State Take 512-th sample
\State Move window by 1 sample\Until{end of sound.}
\end{algorithmic}
\end{algorithm}

\section{Experiments}
\label{sec:exp}

\begin{figure}[htb]

\begin{minipage}[b]{1.0\linewidth}
  \centering
  \centerline{\includegraphics[width=8.5cm]{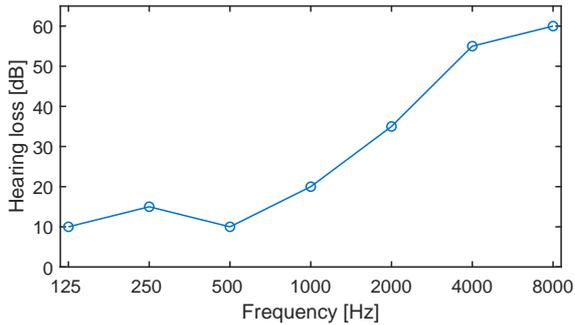}}
\end{minipage}
\caption{Audiogram of the hearing loss that was compensated.}
\label{fig:aud}
\end{figure}

For the experiments we trained the original WaveGlow for normal hearing~\cite{Waveglow2020} and our adapted WaveGlow (IA-TTS) for the better ear of a hearing-impaired listener. His audiogram is shown in Figure~\ref{fig:aud} and is a typical sloping hearing loss. For the loudness calculations, we assumed that the hearing loss is made up of 90\% outer hair cell loss and 10\% inner hair cell loss. A high percentage of outer hair cell loss is typical for a cochlear hearing loss~\cite{Moore2004} and is assumed by algorithms for fitting hearing aids~\cite{Moore2010,Keidser2011}.

We trained both the normal-hearing and adapted IA-TTS model for 500\,000 iterations each with a batch size of 8 on an Nvidia V100 GPU. To evaluate a potential benefit of transfer learning, we trained the IA-TTS model for 125,000 iterations starting from the trained normal-hearing model. The parameters for the WaveGlow architecture and size of Mel spectrograms were kept at the defaults. In particular, WaveGlow consisted of 12 coupling layers and invertible 1x1 convolutions and the Mel spectrograms had 80 bins spanning the frequency range from 0 to 8\,000 Hz. They were based on 1024-point FFTs with a hop size of 256 samples. The learning rate was 0.0001, and the segment length for training was 16\,000 samples. We used the first 10\,000 sounds of the LJSpeech corpus for training and randomly chosen clips between the 11\,000th and 12\,000th clip for evaluations. The code for our adaptations to WaveGlow is available online~\cite{Schlittenlacher2020}.

Individual amplification to compensate for a hearing loss typically increases speech intelligibility because inaudible or barely audible frequency components are restored. The sound quality can be increased because the audibility of the whole spectrum can make sounds more natural or brilliant. However, a complex manipulation of the signal can also introduce audible artefacts that decrease sound quality. For this reason we evaluated both speech quality and speech intelligibility. Listeners rated the sounds on a 5-category scale (1: Bad, 2: Poor, 3: Fair, 4: Good, 5: Excellent) on which mean opinion scores (MOS) were calculated.

\subsection{Evaluation by the target listener}
\label{ssec:restarget}

The listener for whom the sounds were amplified rated the speech quality for seven conditions: (1) The original speech, (2) the original speech linearly amplified by 3.5 dB, (3) the amplified speech, (4) the output of WaveGlow for normal-hearing after 500\,000 iterations, (5) the output of WaveGlow for the amplified speech after 500\,000 iterations, (6) the output of WaveGlow for the amplified speech after 125\,000 iterations, (7) the output of WaveGlow for the amplified speech after 25\,000 iterations of transfer learning. MOS and standard errors of the mean are shown in Table~\ref{tab1}. He rated each condition 18 times, nine times in each of two sessions on two different days.

\begin{table}[htb]
\caption{Speech quality for the target listener: MOS and standard errors of the mean}
\centering
\begin{tabular}{clr}
\toprule
& \multicolumn{1}{c}{Condition} & \multicolumn{1}{c}{MOS} \\
\midrule
1 & Original speech & 4.00 $\pm$ 0.14 \\
2 & Original speech + 3.5 dB & 4.11 $\pm$ 0.11 \\
3 & Individually amplified speech & 4.39 $\pm$ 0.18  \\
4 & Normal WaveGlow, 500K & 3.83 $\pm$ 0.12  \\
5 & Amplified WaveGlow, 500K & 3.33 $\pm$ 0.18  \\
6 & Amplified WaveGlow, 125K & 2.17 $\pm$  0.09 \\
7 & Transfer-learning WaveGlow, 25K & 3.67 $\pm$ 0.16 \\
\bottomrule
\label{tab1}
\end{tabular}
\end{table}

The quality of the individually amplified speech was rated higher than the original speech and slightly linearly amplified speech, with scores of 4.4 compared to 4.0 and 4.1. In contrast to this pattern, the speech generated by WaveGlow that was trained for normal hearing (3.8) was rated similarly to the transfer-learning strategy (3.7) and higher than the IA-TTS that was trained from scratch (3.3).

For ratings of speech intelligibility, all sounds were presented at the same loudness, i.e. mimicking a situation in which the listeners would have adjusted a home assistant to a comfortable loudness. Original speech needed a linear amplification of 3.5 dB to result in the same loudness as individually amplified speech. Ratings for speech intelligibility were obtained for signal-to-noise ratios (SNRs) of 0 and -4 dB relative to original speech (Table~\ref{tabsi}). The noise was steady and had the average spectrum of speech.

\begin{table}[htb]
\caption{Speech intelligibility for the target listener: MOS and standard errors of the mean}
\centering
\begin{tabular}{lrr}
\toprule
\multicolumn{1}{c}{Condition} & \multicolumn{1}{c}{0 dB SNR} & \multicolumn{1}{c}{-4 dB SNR}\\
\midrule
Original speech & 3.67 $\pm$ 0.16 & 1.89 $\pm$ 0.19 \\
IA speech & 4.89 $\pm$ 0.10 & 3.78 $\pm$ 0.14 \\
Normal WaveGlow, 500K & 3.56 $\pm$ 0.17 & 2.11 $\pm$ 0.25 \\
Amplified WaveGlow, 500K & 4.56 $\pm$ 0.17 & 3.11 $\pm$ 0.25 \\
TL WaveGlow, 25K & 4.78 $\pm$ 0.14 & 3.33 $\pm$ 0.22 \\

\bottomrule
\label{tabsi}
\end{tabular}
\end{table}

At a given noise level, intelligibility was considerably better for individually amplified speech than for normal speech at the same loudness. Furthermore, the transfer-learning approach reached about the same intelligibility as its targets and the IA-TTS that was trained from scratch.

\begin{figure}[htb]

\begin{minipage}[b]{1.0\linewidth}
  \centering
  \centerline{\includegraphics[width=8.5cm]{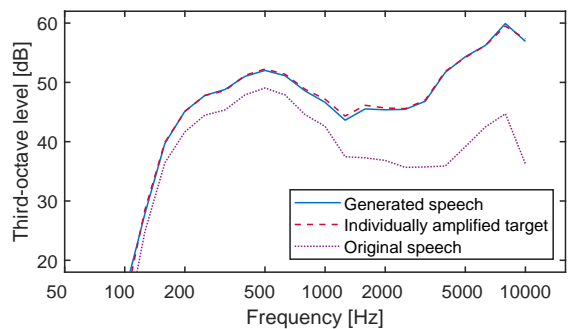}}
\end{minipage}
\caption{Average third-octave spectra of the test set for the speech produced by transfer learning after 25\,000 iterations (solid blue line), the individually amplified targets (red dashed line) and the original speech (dotted purple line).}
\label{fig:tl}
\end{figure}

The transfer-learning approach adjusted the spectra of the produced speech quickly to those of the individually amplified targets. Figure~\ref{fig:tl} shows that the average third-octave spectrum after 25\,000 iterations of transfer learning (solid blue line) matches that of the individually amplified targets (dashed red line), both in shape and overall level. This indicates that possibly even fewer iterations would have been sufficient to change the spectrum from that of original speech to the new target.

\subsection{Evaluation by normal-hearing listeners}
\label{ssec:resnh}

Normal-hearing listeners evaluated the WaveGlow model after 60\,000, 125\,000, 250\,000 and 500\,000 iterations, and a publicly available model~\cite{Waveglow2020} trained on 580\,000 iterations and a batch size of 24 to test how many iterations are necessary to reach a sufficient quality. The mean opinion scores for speech quality were compared to STOI values~\cite{Taal2010} to evaluate how well they can be predicted by an objective metric.

Furthermore, the listeners rated the sounds after the inverse amplification was applied to individually amplified sounds. Artefacts due to windowing and the non-linearity of the auditory-filter based amplification/attenuation would possibly occur twice using this strategy rather than canceling out. Thus, the scores give a pessimistic estimate of the decrease in speech quality due to artefacts and glitches that are caused by the algorithm. The inverse amplification was aimed at restoring the loudness for normal-hearing listeners for sounds that were amplified to compensate for a hearing loss, i.e. was the same as Algorithm~\ref{alg1} except that the logic for steps 2 and 3 of the gain calculation was reversed.

The listeners also rated the quality of speech that was generated from text using a pre-trained Tacotron2~\cite{Tacotron2020} and WaveGlow trained on 500\,000 iterations. These sentences were not part of LJSpeech. 49 ratings were collected for each condition. MOS, standard errors of the mean and STOI values are shown in Table~\ref{tab2}. 

\begin{table}[htb]
\caption{MOS and standard errors of the mean for normal-hearing listeners and STOI relative to original speech}
\centering
\begin{tabular}{lrr}
\toprule
\multicolumn{1}{c}{Condition} & \multicolumn{1}{c}{MOS} & \multicolumn{1}{c}{STOI} \\
\midrule
Original speech & 4.55 $\pm$ 0.10 & 1 \\
WaveGlow, 60K & 1.92 $\pm$ 0.13 & 0.87 \\
WaveGlow, 125K & 2.59 $\pm$ 0.12 & 0.92 \\
WaveGlow, 250K & 2.49 $\pm$ 0.12 & 0.92 \\
WaveGlow, 500K & 4.04 $\pm$ 0.12 & 0.97 \\
WaveGlow, 580K, batch size 24 & 4.14 $\pm$ 0.13 & 0.98 \\
Amplified WaveGlow, 500K + & \multirow{2}{*}{3.51 $\pm$ 0.13} &  \multirow{2}{*}{0.94} \\
inverse amplification & & \\
Tacotron2 + WaveGlow & 3.71 $\pm$ 0.14 & - \\
Tacotron2 + amplif. WaveGlow + & \multirow{2}{*}{3.20 $\pm$ 0.14} &  \multirow{2}{*}{-} \\
inverse amplification & & \\

\bottomrule
\label{tab2}
\end{tabular}
\end{table}

MOS and STOI correlate highly, r(5)=0.97, p$<$0.001. Both captured the momentary break-in during training at 250\,000 iterations, which was also reflected by the training error (not shown here). The MOS for the IA-TTS that was followed by the inverse amplification was predicted accurately by STOI. Applying the individual amplification and its inverse decreased the speech quality by about 0.5 categories compared to the equivalent conditions without amplification. Generating speech from text rather than from Mel spectrograms decreased the ratings by about 0.3 categories.

\section{DISCUSSION}
\label{sec:discussion}

In this paper we presented a TTS setup for the hearing impaired that employs an algorithm to compensate for the hearing loss offline and thus can be complex. The proposed amplification algorithm builds on the principles of hearing-aid processing but its hop size of only one sample leads to a restoration of loudness at a high resolution not only in the frequency domain, but also in the time and input level domains. It has been evaluated by only one hearing-impaired listener so far but the high speech-quality MOS of 4.39 for individually amplified speech, which is higher than his rating for original speech and close to the MOS of normal-hearing listeners for original speech, is promising. As expected, the speech intelligibility in a given noise is much higher with the individual amplification at the same loudness.

The normal-hearing listeners reproduced MOS for original speech (4.55) and the maximally trained neural vocoder (4.14) from the literature~\cite{Oord2016, Prenger2019}. They rated the sounds for which the manipulation was applied twice about 0.5 categories worse than their unmodified counterparts, which may be an upper limit for the non-linear distortions that the individual amplification introduces, and is in line with the ratings of the hearing-impaired listener for the fully trained WaveGlow conditions.

The hearing-impaired listener rated the individually amplified sounds at about the same quality as the original sounds: slightly better when applied directly to the sounds but slightly worse when training the IA-TTS from scratch, and about equally good for the transfer-learning approach. Speech intelligibility was systematically higher for all individually amplified sounds when they had the same loudness as original speech.

The transfer-learning approach led to a quick adaptation of a pretrained model to a given hearing loss, reaching the same spectra and the increased speech intelligibility of individual amplification in 25\,000 iterations or less and showing about the same speech quality as original speech. The results indicate that it performed even slightly better than the IA-TTS model that was trained from scratch. This may be due to difficulties of training a TTS system to waveform targets with more high-frequency content. In a next step, we are going to validate the IA-TTS for more types of hearing losses and listeners, and vary the number of iterations in the transfer-learning approach in listening tests.


The amplification algorithm could be optimized to further reduce distortions, especially in the high-frequency region for which the analysis window of 1024 samples (46 ms) was relatively long. Further improvement may be obtained by making the individual amplification reproduce the time-varying specific loudness of speech rather than that of a broadband noise with the same auditory-filter levels. However, this is computationally expensive for a large dataset since it does not allow the use of look-up tables. These optimization procedures can be guided by STOI, which showed a high correlation to the scores including for the condition in which the algorithm was applied twice.

At inference time, the IA-TTS model has the same complexity as that for conventional TTS. However, it needs to be trained for an individual hearing loss. The spectra produced by the IA-TTS were much closer to the individually amplified targets than to the original speech already after 25\,000 iterations of transfer learning, and the score for sound quality was in the same region as those for fully trained TTS systems. The computation time for an individual amplification can further be decreased by processing less material for training. The transfer-learning approach for enhancing TTS in noise using the Lombard effect produced good results with only 2.5 hours of training material. A further option is to train the IA-TTS for a set of audiogram types~\cite{Bisgaard2010} so that a reasonably close amplification is available without or before individual optimization.

The presented IA-TTS setup is modular, and systems other than WaveGlow and Tacotron2 may be chosen. The subjective evaluations of the present study suggest that the individual amplification to compensate for a hearing loss produces high-quality audio similar to that of unprocessed speech, results in high speech intelligibility in noise without increasing loudness, and that a pre-trained TTS model can be adjusted to a new hearing loss with relatively few iterations of training.

\bibliographystyle{IEEEtran}
\bibliography{refs}


\end{document}